# Steganography of VoIP Streams

Wojciech Mazurczyk, Krzysztof Szczypiorski

Warsaw University of Technology,
Faculty of Electronics and Information Technology,
Institute of Telecommunications,
15/19 Nowowiejska Str., 00-665 Warsaw, Poland
{W.Mazurczyk, K.Szczypiorski}@tele.pw.edu.pl

**Abstract.** The paper concerns available steganographic techniques that can be used for creating covert channels for VoIP (Voice over Internet Protocol) streams. Apart from characterizing existing steganographic methods we provide new insights by presenting two new techniques. The first one is network steganography solution which exploits free/unused protocols' fields and is known for IP, UDP or TCP protocols but has never been applied to RTP (Real-Time Transport Protocol) and RTCP (Real-Time Control Protocol) which are characteristic for VoIP. The second method, called LACK (Lost Audio Packets Steganography), provides hybrid storage-timing covert channel by utilizing delayed audio packets. The results of the experiment, that was performed to estimate a total amount of data that can be covertly transferred during typical VoIP conversation phase, regardless of steganalysis, are also included in this paper.

## 1 Introduction

VoIP is one of the most popular services in IP networks and it stormed into the telecom market and changed it entirely. As it is used worldwide more and more willingly, the traffic volume that it generates is still increasing. That is why VoIP is suitable to enable hidden communication throughout IP networks. Applications of the VoIP covert channels differ as they can pose a threat to the network communication or may be used to improve the functioning of VoIP (e.g. security like in [12] or quality of service like in [13]). The first application of the covert channel is more dangerous as it may lead to the confidential information leakage. It is hard to assess what bandwidth of covert channel poses a serious threat – it depends on the security policy that is implemented in the network. For example, US Department of Defense specifies in [24] that any covert channel with bandwidth higher than 100 bps must be considered insecure for average security requirements. Moreover for high security requirements it should not exceed 1 bps.

In this paper we present available covert channels that may be applied for VoIP during conversation phase. A detailed review of steganographic methods that may be applied during signalling phase of the call can be found in [14]. Here, we introduce two new steganographic methods that, to our best knowledge, were not described earlier. Next, for each of these methods we estimate potential bandwidth to evaluate experimentally how much information may be transferred in the typical IP telephony call.

The paper is organized as follows. In Section 2 we circumscribe the VoIP traffic and the communication flow. In Section 3, we describe available steganographic methods that can be utilized to create covert channels in VoIP streams. Then in Section 4 we present results of the experiment that was performed. Finally, Section 5 concludes our work.

## 2 VoIP Communication Flow

VoIP is a real-time service that enables voice conversations through IP networks. It is possible to offer IP telephony due to four main groups of protocols:

a. *Signalling protocols* that allow to create, modify, and terminate connections between the calling parties – currently the most popular are SIP [18], H.323 [8], and H.248/Megaco [4],
b. *Transport protocols* – the most important is RTP [19], which provides end-to-end network transport functions suitable for applications transmitting real-time audio. RTP is usually used in conjunction with UDP (or rarely TCP) for transport of digital voice stream,
c. *Speech codecs* e.g. G.711, G.729, G.723.1 that allow to compress/decompress digitalized human voice and prepare it for transmitting in IP networks.
d. Other *supplementary protocols* like RTCP [19], SDP, or RSVP etc. that complete VoIP functionality. For purposes of this paper we explain the role of RTCP protocol: RTCP is a control protocol for RTP and it is designed to monitor the Quality of Service parameters and to convey information about the participants in an ongoing session.

Generally, IP telephony connection consists of two phases: a *signalling phase* and a *conversation phase*. In both phases certain types of traffic are exchanged between calling parties. In this paper we present a scenario with SIP as a signalling protocol and RTP (with RTCP as control protocol) for audio stream transport. That means that during the signalling phase of the call certain SIP messages are exchanged between SIP endpoints (called: SIP User Agents). SIP messages usually traverse through SIP network servers: proxies or redirects that allow end-users to locate and reach each other. After this phase, the conversation phase begins, where audio (RTP) streams flow bi-directly between a caller and a callee. VoIP traffic flow described above and distinguished phases of the call are presented in Fig. 1. For more clarity we omitted the SIP network server in this diagram. Also potential security mechanisms in traffic exchanges were ignored.

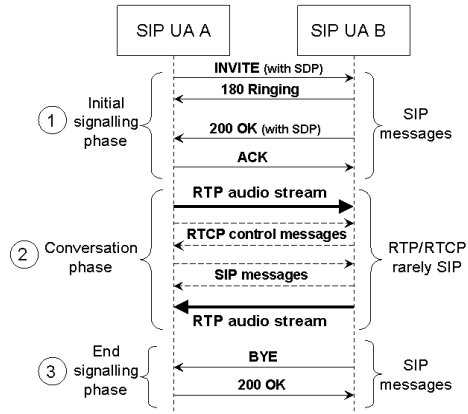

**Fig. 1.** VoIP call setup based on SIP/SDP/RTP/RTCP protocols (based on [9])

## 3 Covert Channels in VoIP Streams Overview and New Insights

Besides characterizing IP telephony traffic flow Fig. 1 also illustrates steganographic model used in this paper for VoIP steganography evaluation. The proposed model is as follows. Two users *A* and *B* are performing VoIP conversation while simultaneously utilizing it to send steganograms by means of all possible steganographic methods that can be applied to IP telephony protocols. We assume that both users control their end-points (transmitting and receiving equipment) thus they are able to modify and inspect the packets that are generated and received. After modifications at calling endpoint, packets are transmitted through communication channel which may introduce negative effects e.g. delays, packet losses or jitter. Moreover, while traveling through network packets can be inspected and modified by an active warden [5]. Active wardens act like a semantic and syntax proxy between communication sides. They are able to modify and normalize exchanged traffic in such a way that it does not break, disrupt or limit any legal network communication or its functionality. Thus, active wardens can inspect all the packets sent and modify them slightly during the VoIP call. It must be emphasized however that they may not erase or alter data that can be potentially useful for VoIP non-steganographic (overt) users. This assumption forms important active wardens' rule although sometimes elimination of the covert channel due to this rule may be difficult.

To later, in section 4, practically evaluate covert channels that can be used for VoIP transmission we must first define three important measures that characterizes them and which must be taken into consideration during VoIP streams covert channels analysis. These measures are:
- *Bandwidth* that may be characterized with *RBR* (Raw Bit Rate) that describes how many bits may be sent during one time unit [bps] with the use of all steganographic techniques applied to VoIP stream (with no overheads included) or *PRBR* (Packet Raw Bit Rate) that circumscribe how much information may be covertly sent in one packet [bits/packet],
- *Total amount of covert data* [bits] transferred during the call that may be sent in one direction with the use of all applied covert channels methods for typical VoIP call. It means that, regardless of steganalysis, we want to know how much covert information can be sent during typical VoIP call,
- *Covert data flow distribution during the call* – how much data has been transferred in a certain moment of the call.

We will be referencing to abovementioned measures during the following sections while presenting available steganographic methods for VoIP communication and later during the experiment description and results characterization.

In this section we will provide an overview of existing steganographic techniques used for creation of covert channels in VoIP streams and present new solutions. As described earlier during the conversation phase audio (RTP) streams are exchanged in both directions and additionally, RTCP messages may be sent. That is why the available steganographic techniques for this phase of the call include:
- *IP/UDP/TCP/RTP* protocols steganography in network and transport layer of TCP/IP stack,
- *RTCP* protocol steganography in application layer of TCP/IP stack,
- *Audio watermarking* (e.g. LSB, QIM, DSSS, FHSS, Echo hiding) in application layer of TCP/IP stack,
- *Codec SID frames* steganography in application layer of TCP/IP stack,
- *Intentionally delayed audio packets* steganography in application layer of TCP/IP stack,
- *Medium dependent* steganographic techniques like HICCUPS [22] for VoWLAN (Voice over Wireless LAN) specific environment in data link layer of TCP/IP stack.

Our contribution in the field of VoIP steganography includes the following:
- Describing RTP/RTCP protocols' fields that can be potentially utilized for hidden communication,
- Proposing *security mechanisms fields steganography* for RTP/RTCP protocols,
- Proposing intentionally delayed audio packets steganographic method called LACK (Lost Audio Packets Steganographic Method).

### 3.1 IP/TCP/UDP Protocols Steganography

TCP/UDP/IP protocols steganography utilizes the fact that only few fields of headers in the packet are changed during the communication process ([15], [1], [17]). Covert data is usually inserted into redundant fields (provided, but often unneeded) for abovementioned protocols and then transferred to the receiving side. In TCP/IP stack, there is a number of methods available, whereby covert channels can be established and data can be exchanged between communication parties secretly. An analysis of the headers of TCP/IP protocols e.g. IP, UDP, TCP results in fields that are either unused or optional [15], [25]. This reveals many possibilities where data may be hidden and transmitted. As described in [15] the IP header possesses fields that are available to be used as covert channels. Notice, that this steganographic method plays an important role for VoIP communication because protocols mentioned above are present in every packet (regardless, if it is a signalling message, audio packet, or control message). For this type of steganographic method as well as for other protocols in this paper (RTP and RTCP steganography) achieved steganographic bandwidth can be expressed as follows:

$$PRBR_{NS} = \frac{\left(SB_0 + \sum_{j=1}^{l} SB_j\right)}{l+1} \quad [bits/packet] \qquad (1)$$

where:

$PRBR_{NS}$ (Packet Raw Bit Rate) denotes bandwidth of the covert channel created by IP/TCP/UDP steganography [bits/packet],

$SB_0$ is total amount of bits for IP/TCP/UDP protocols that can be covertly send in the fields of the first packet. This value differs from the value achieved for the following packets because in the first packet initial values of certain fields can be used (e.g. sequence number for TCP protocol),

$SB_j$ denotes total amount of bits for IP/TCP/UDP protocols that can be covertly sent in the fields of the following packets,

$l$ is number of packets send besides first packet.

### 3.2 RTP Protocols Steganography
### 3.2.1 RTP Free/Unused Fields Steganography

In conversation phase of the call when the voice stream is transmitted, besides protocols presented in section 3.1 also the fields of RTP protocol may be used as a covert channel. Fig. 5 presents the RTP header.

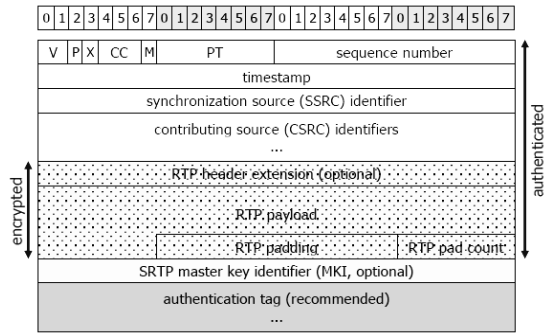

**Fig. 5.** RTP header with marked sections that are encrypted and authenticated

RTP provides the following opportunities for covert communication:
- *Padding* field may be needed by some encryption algorithms. If the padding bit (P) is set, the packet contains one or more additional padding octets at the end of header which are not a part of the payload. The number of the data that can be added after the header is defined in the last octet of the padding as it contains a count of how many padding octets should be ignored, including itself,
- *Extension header* (when X bit is set) – similar situation as with the padding mechanism, a variable-length header extension may be used,
- Initial values of the *Sequence Number* and *Timestamp* fields – because both initial values of these fields must be random, the first RTP packet of the audio stream may be utilized for covert communication,
- Least significant bits of the *Timestamp* field can be utilized in a similar way as proposed in [6].

It must be emphasized however that steganography based on free/unused/optional fields for RTP protocol (as well as for protocols mentioned in section 3.1) may be potentially eliminated or greatly limited by active wardens. Normalization of RTP headers' fields values (e.g. applied to *Timestamps*) or small modifications applied may be enough to limit covert bandwidth. On the other hand it is worth noting that so far no documented active warden implementation exists.

### 3.2.2 RTP Security Mechanisms Steganography

There is also another way to create high-bandwidth covert channel for RTP protocol. In Fig. 5 one can see what parts of RTP packet is secured by using encryption (payload and optionally header extension if used) and authentication (*authentication tag*). For steganographic purposes we may utilize security mechanisms' fields. The main idea is to use *authentication tag* to transfer data in a covert manner. In SRTP (Secure RTP) standard [2] it is recommended that this field should be 80 bits long but lower values are also acceptable (e.g. 32 bits). Similar steganographic method that utilizes security mechanism fields was proposed for e.g. IPv6 in [11]. By altering content of fields like *authentication tag* with steganographic data it is possible to create covert channel because data in these fields is almost random due to the cryptographic mechanism operations. That is why it is hard to detect whether they carry real security data or hidden information. Only receiving calling party, as he is in possession of pre-shared key (*auth_key*) is able to determine that. For overt users wrong authentication data in packet will mean dropping it. But because receiving user is controlling its VoIP equipment, when *authentication tag* fields are utilized as covert channel, he is still able to extract steganograms in spite of invalid authentication result.

Thus, most of steganalysis methods will fail to uncover this type of secret communication. The only solution is to strip off/erase such fields from the packets but this is a serious limitation for providing security services for overt users. Moreover it will be violation of the active warden rule (that no protocol's semantic or syntax will be disrupted).

Because the number of RTP packets per one second is rather high (depends on the voice frame generation interval) exploiting this tag provides a covert channel that bandwidth can be expressed as follows:

$$RBR_{SRTP} = SB_{AT} \cdot \frac{1000}{I_p} \quad [bits/s] \tag{2}$$

where:

$RBR_{SRTP}$ (Raw Bit Rate) denotes bandwidth of the covert channel created by RTP security mechanism steganography (in bits/s),

$SB_{AT}$ is total amount of bits in *authentication tag* for SRTP protocol (typically 80 or 32 bits),

$I_p$ describes voice packet generation interval, in miliseconds (typically from 10 to 60 ms).

For example, consider a scenario in which *authentication tag* is 32 bits long and audio packet is generated each 20 ms. Based on equation 2 we can calculate that $RBR_{SRTP}$ = 1.6 kbit/s which is considerably high result for bandwidths of covert channel presented in this paper.

### 3.3 RTCP Protocol Steganography
### 3.3.1 RCTP Free/Unused Fields Steganography

To our best knowledge this is the first proposal to use RTCP protocol messages as a covert channel. RTCP exchange is based on the periodic transmission of control packets to all participants in the session. Generally it operates on two types of packets (reports) called: Receiver Report (RR) and Sender Report (SR). Certain parameters that are enclosed inside those reports may be used to estimate network status. Moreover all RTCP messages must be sent in compound packet that consists of at least two individual types of RTCP reports. Fig. 6 presents headers of SR and RR reports of the RTCP protocol.

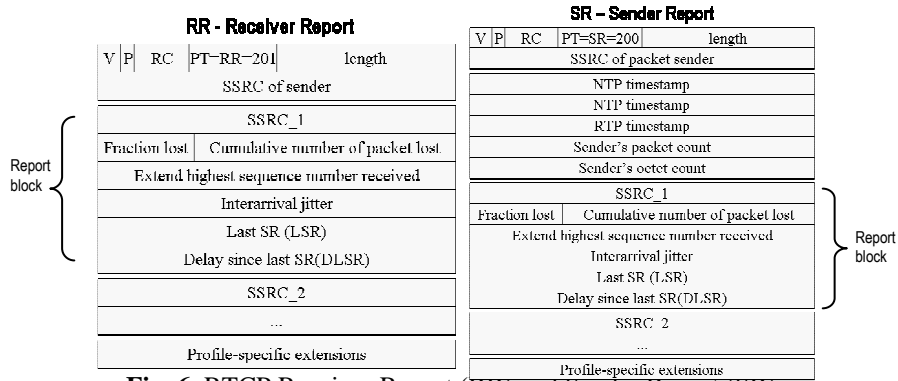

**Fig. 6.** RTCP Receiver Report (RR) and Sender Report (SR)

For sessions with small number of the participants the interval between the RTCP messages is 5 seconds and moreover sending RTCP communication (with overhead) should not exceed 5% of the session's available bandwidth. For creating covert channels report blocks in SR and RR reports (marked in Fig. 6) may be utilized. Values of the parameters transferred inside those reports (besides SSRC_1 which is the source ID) may be altered, so the amount of information that may be transferred in each packet is 160 bits. It is clear, that if we use this type of steganographic technique, we lose some (or all) of RTCP functionality (it is a cost to use this solution). Other free/unused fields in these reports may be also used in the similar way. For example *NTP Timestamp* may be utilized in a similar way as proposed in [6].

Other RTCP packet types include: SDES, APP or BYE. They can also be used in the same way as SR and RR reports. So the total PRBR for this steganographic technique is as follows:

$$PRBR_{RTCP} = S_{CP} \cdot N_{RB} \cdot S_{RB} \ [bits/packet] \quad (3)$$

where:
    $PRBR_{RTCP}$ (Packet Raw Bit Rate) denotes bandwidth of the covert channel created with RCTP Free/Unused Fields Steganography (in bits/packet),
    $S_{CP}$ denotes size of the compound RTCP packet (the number of RTCP packet types inside the compound one),
    $N_{RB}$ is number of report blocks inside each RTCP packet type,
    $S_{RB}$ is the number of bits that can be covertly send in one RTCP report block.

It is also worth noting that RTCP messages are based on IP/UDP protocols, so additionally, for one RTCP packet, both protocols can be used for covert transmission.

To improve capacity of this covert channel one may send RTCP packets more frequently then each 5 seconds (which is default value proposed in standard) although it will be easier to uncover. Steganalysis of this method is not so straightforward as in case of security mechanism fields steganography. Active warden can be used to eliminate or greatly limit the fields in which hidden communication can take place although it will be serious limitation of RTCP functionality for overt users.

### 3.3.2 RTCP Security Mechanisms Steganography

Analogously as for RTP protocol the same steganographic method that uses SRTP security mechanism may be utilized for RTCP and the achieved $RBR_{RTCP}$ rate is as follows:

$$RBR_{SRTCP} = \frac{SB_{AT} \cdot l}{T} \ [bits/s] \quad (4)$$

where:
    $RBR_{SRTCP}$ (Raw Bit Rate) denotes bandwidth of the covert channel created with SRTP security mechanism steganography [in bps],
    $SB_{AT}$ is total amount of bits in *authentication tag* for SRTP protocol,
    $T$ denotes duration of the call (in seconds),
    $l$ is number of RTCP messages exchanged during the call of length $T$.

### 3.4 Audio Watermarking

The primary application of audio watermarking was to preserve copyrights and/or intellectual properties called DRM (Digital Right Management). However, this technique can be also used to create effective covert channel inside a digital content. Currently there is a number of audio watermarking algorithms available. The most popular methods that can be utilized in real-time communication for VoIP service, include: *LSB* (Least Significant Bit), *QIM* (Quantization Index Modulation), *Echo Hiding*, *DSSS* (Direct Sequence Spread Spectrum), and *FHSS* (Frequency Hopping Spread Spectrum) [3]. For these algorithms the bandwidth of available covert channels depends mainly on the sampling rate and the type of audio material being encoded.

Moreover, if covert data rate is too high it may cause voice quality deterioration and increased risk of detection. In Table 1 examples of digital watermarking data rates are presented under conditions that they do not excessively affect quality of the conversation and limit probability of disclosure. Based on those results one can clearly see that, besides *LSB* watermarking, other audio watermarking algorithms covert channels' bandwidth range from few to tens bits per second.

**Table 1.** Audio watermarking algorithms and their experimentally calculated RBRs

| Audio watermarking algorithm | Covert bandwidth RBR (based on [21]) | Covert bandwidth RBR (based on [1]) |
|---|---|---|
| LSB | 1 kbps / 1 kHz (of sampling rate) | 4 kbps |
| DSSS | 4 bps | 22.5 bps |
| FHSS | - | 20.2 bps |
| Echo Hiding | 16 bps | 22.3 bps |

Thus, we must consider that each audio watermarking algorithm affects perceived quality of the call. That means that there is a necessary tradeoff between the amount of data to be embedded and the degradation in users' conversation. On the other hand by using audio watermarking techniques we gain an effective steganographic method: because of the audio stream flow the achieved bandwidth of the covert channel is constant. Thus, although the bit rate of audio watermarking algorithms is usually not very high, it still may play important role for VoIP streams covert channels.

Steganalysis of audio watermarking methods (besides for LSB algorithm which is easy to eliminate) is rather difficult and must be adjusted to watermarking algorithm used. It must be emphasized however that if hidden data embedding rate is chosen reasonably then detecting of the audio watermarking is hard but possible and in most cases erasing steganogram means great deterioration of voice quality.

### 3.5 Speech Codec Silence Insertion Description (SID) Frames Steganography

Speech codecs may have built-in or implement mechanisms like Discontinuous Transmission (DTX)/VAD (Voice Activity Detection)/CNG (Comfort Noise Generation) for network resources (e.g. bandwidth) savings. Such mechanisms are able to determine if voice is present in the input signal. If it is present, voice would be coded with the speech codec in other case, only a special frame called Silence Insertion Description (SID) is sent. If there is a silence, instead of sending large voice packets that do not contain conversation data only small amount of bits are transmitted. Moreover, during silence periods, SID frames may not be transferred periodically, but only when the background noise level changes. The size of this frame depends on the speech codec used e.g. for G.729AB it is 10 bits per frame while for G.723.1 it is 24 bits per frame. Thus, when DTX/VAD/CNG is utilized, during the silence periods SID frames can be used to covertly transfer data by altering information of background noise with steganogram. In this case no new packets are generated and the covert bandwidth depends on the speech codec used. It is also possible to provide higher bandwidth of the covert channel by influencing rate at which SID frames are issued. In general, the more of these frames are sent the higher the bandwidth of the covert channel. It must be however noted that the covert bandwidth for this steganographic is rather low.

What is important, for this steganographic method steganalysis is simple to perform. Active warden that is able to modify some of the bits in SID frames (e.g. least significant) can eliminate or greatly reduce the bandwidth of this method.

**3.6 LACK: Intentionally Delayed Audio Packets Steganography**

To our best knowledge this is the first proposal of using intentionally delayed (and in consequence lost) packets payloads as a covert channel for VoIP service. Although there was an attempt how to use channel erasures at the sender side for covert communication [20] but this solution characterizes low bandwidth especially if we use it for VoIP connection (where the packet loss value must be limited). It is natural for IP networks that some packets can be lost due to e.g. congestion. For IP telephony, we consider a packet lost when:
- It does not reach the destination point,
- It is delayed excessive amount of time (so it is no longer valid), and that is why it may not be used for current voice reconstruction in the receiver at the arrival time.

Thus, for VoIP service when highly delayed packet reaches the receiver it is recognized as lost and then discarded. We can use this feature to create new steganographic technique. We called this method LACK (Lost Audio Packets Steganographic Method). In general, the method is intended for a broad class of multimedia, real-time applications. The proposed method utilizes the fact that for usual multimedia communication protocols like RTP excessively delayed packets are not used for reconstruction of transmitted data at the receiver (the packets are considered useless and discarded). The main idea of LACK is as follows: at the transmitter, some selected audio packets are intentionally delayed before transmitting. If the delay of such packets at the receiver is considered excessive, the packets are discarded by a receiver not aware of the steganographic procedure. The payload of the intentionally delayed packets is used to transmit secret information to receivers aware of the procedure. For unaware receivers the hidden data is "invisible".

Thus, if we are able to add enough delay to the certain packets at the transmitter side they will not be used for conversation reconstruction. Because we are using legitimate VoIP packets we must realize that in this way we may influence conversation quality. That is why we must consider the accepted level of packet loss for IP telephony and do not exceed it. This parameter is different for various speech codecs as researched in [16] e.g. 1% for G.723.1, 2% for G.729A, 3% for G.711 (if no additional mechanism is used to cope with this problem) or even up to 5% if mechanisms like PLC (Packet Loss Concealment) is used. So the number of packets that can be utilized for proposed steganographic method is limited. If we exceed packet loss threshold for chosen codec then there will be significant decrease in voice quality.

Let us consider RTP (audio) stream ($S$) that consists of $n$ packets ($a_n$):

$$S = (a_1, a_2, a_3, \ldots, a_n) \text{ and } n = T / I_f \qquad (5)$$

where:
  $S$ denotes RTP (audio) stream,
  $a_n$ is n-th packet in the audio stream $S$,
  $n$ a number of packets in audio stream.

For every packet ($a_n$) at the transmitter output total delay ($d_T$) is equal to:

$$d_T(a_n) = \sum_{m=1}^{3} d_m \qquad (6)$$

where:
  $d_1$ is speech codec processing delay,
  $d_2$ is codec algorithm delay,
  $d_3$ is packetization delay.

Now, from stream $S$ we choose $i$-th packet $a_i$ with a probability ($p_i$):

$$p_i < p_{Lmax} \qquad (7)$$

where:
  $p_{Lmax} \in \{1\%, 5\%\}$ where 1% packet loss ratio is for VoIP without PLC mechanism and 5% packet loss ratio is for VoIP with PLC mechanism.

To be sure that the RTP packet will be recognized as lost at the receiver, as mentioned earlier, we have to delay it by certain value. For the proposed steganographic method two important parameters must be considered and set to the right value: amount of time by which the chosen packet is delayed ($d_4$), to ensure that it will be considered as lost at the receiver side and the packet loss probability ($p_i$) for this steganographic method, to ensure that in combination with $p_{Lmax}$ probability will not degrade perceived quality of the conversation. To properly choose a delay value, we must consider capacity of the receiver's de-jitter buffer. The de-jitter buffer is used to alleviate the jitter effect (variations in packets arrival time caused by queuing, contention and serialization in the network). Its value (usually between 30-70 ms) is important for the end-to-end delay budget (which should not exceed 150 ms). That is why we must add $d_4$ delay (de-jitter buffer delay) to the $d_T$ value for the chosen packet ($a_i$). If we ensure that $d_4$ value is equal or greater than de-jitter buffer delay at the receiver side the packet will be considered lost. So the total delay ($d_T$) for $a_i$ packets with additional $d_4$ delay looks as follows (8):

$$d_T(a_i) = \sum_{m=1}^{4} d_m \qquad (8)$$

where $d_4$ is de-jitter buffer delay.

Now that we are certain that the chosen packet ($a_i$) is considered lost at the receiver, we can use this packet's payload as a covert channel.

As mentioned earlier, the second important measure for proposed steganographic method is a probability $p_i$. To properly calculate its value we must consider the following simplified packet loss model:

$$p_T = 1 - (1 - p_N)(1 - p_i) \qquad (9)$$

where:
  $p_T$ denotes total packet loss probability in the IP network that offers VoIP service with the utilizing of delayed audio packets,

$p_N$ is a probability of packet loss in the IP network that offers VoIP service without the utilizing delayed audio packets (network packet loss probability),

$p_i$ denotes a maximum probability of the packet loss for delayed audio packets.

When we transform (9) to calculate $p_i$ we obtain:

$$p_i \leq \frac{p_T - p_N}{1 - p_N} \qquad (10)$$

From (10) one can see that probability $p_i$ must be adjusted to the network conditions. Information about network packet loss probability may be gained e.g. from the RTCP reports during the transmission. So, based on earlier description, we gain a covert channel with *PRBR* (Packet Raw Bit Rate) that can be expressed as follows:

$$PRBR = r \cdot \frac{I_f}{1000} \cdot p_i \quad [bits/packet] \qquad (11)$$

where *r* is the speech codec rate.

And available bandwidth expressed in *RBR* (Raw Bit Rate) can be described with the following equation (12):

$$RBR = r \cdot p_i \quad [bits/s] \qquad (12)$$

For example, consider a scenario with G.711 speech codec where: speech codec rate: $r$ = 64 kbit/s and $p_i$ = 0.5% and $I_f$ = 20 ms. For these exemplary values RBR is 320 b/s and PRBR is 6.4 bits/packet. One can see that the available bandwidth of this covert channel is proportional to the speech codec frame rate, the higher the rate, the higher the bandwidth. So the total amount of information ($I_T$) that can be covertly transmitted during the call of length *d* (in seconds) is:

$$I_T = d \cdot RBR = d \cdot r \cdot p \quad [bits] \qquad (13)$$

Proposed steganographic method has certain advantages. Most of all, although it is an application layer steganography technique, it is less complex than e.g. most audio steganography algorithms and the achieved bandwidth is comparable or even higher.

Steganalysis of LACK is harder than in case of other steganographic methods that are presented in this paper. This is mainly because it is common for IP networks to introduce losses. If the amount of the lost packets used for LACK is kept reasonable then it may be difficult to uncover hidden communication. Potential steganalysis methods include:

- Statistical analysis of the lost packets for calls in certain network. This may be done by passive warden (or other network node) e.g. based on RTCP reports (Cumulative number of packets lost field) or by observing RTP streams flow (packets' sequence numbers). If for some of the observed calls the number of lost packets is higher than it can indicate potential usage of the LACK method,
- Active warden which analyses all RTP streams in the network. Based on the SSRC identifier and fields: S*equence number* and *Timestamp* from RTP header it can identify packets that are already too late to be used for voice reconstruction. Then active warden may erase their payloads fields or simply drop them. One problem with this steganalysis method is how greatly the packets' identifying numbers must differ from other packets in the stream to be discarded without eliminating really delayed packets that may be still used for conversation. The size of jitter buffer at the receiver is not fixed (and may be not constant) and its size is unknown to active warden. If active warden drops all delayed packets then it could remove packets that still will be usable for voice reconstruction. In effect, due to active warden operations quality of conversation may deteriorate.

Further in-depth steganalysis for LACK is surely required and is considered as future work.

### 3.7 Medium Dependent Steganography

Medium dependent steganography typically uses layer 1 or layer 2 of ISO OSI RM. For VoIP e.g. in homogenous WLAN environment data link layer methods that depend on available medium like HICCUPS [22] system can be utilized. Exemplary, the data rate for this system is 216 kbit/s (IEEE 802.11g 54 Mbit/s, changing of frame error rate from 1.5% into 2.5%, bandwidth usage 40%).

It must be emphasized however that this steganographic method is difficult to implement as it require modification to network cards. Moreover, steganalysis for HICCUPS is difficult too as it necessary to analyze frames in physical layer of OSI RM model.

## 4 Experimental Evaluation of VoIP Streams Covert Channels Bandwidth

Total achieved covert channel bandwidth ($B_T$) for the whole VoIP transmission is a sum of each, particular bandwidth of each steganographic methods that are used during voice transmission (each steganographic subchannel). It can be expressed as follows:

$$B_T = \sum_{j=1}^{k} B_j \quad (14)$$

where:
$B_T$ denotes a total bandwidth for the whole VoIP voice transmission (may be expressed in RBR or PRBR),
$B_j$ describes a bandwidth of the covert channel created by each steganographic method used during VoIP call (may be expressed in RBR or PRBR),
$k$ is a number of steganographic techniques used for VoIP call.

The value of $B_T$ is not constant and depends on the following factors:
- The number of *steganographic techniques* applied to the VoIP call,
- The *choice of the speech codec* used. Three important aspects must be considered here: *compression rate* (e.g. G.711 achieves 64 kbit/s while G729AB only 8 kbit/s), *size of the voice frame* that is inserted into each packet and *voice packet generation interval*. Compression rate influences the available bandwidth of the steganographic methods that relay on it. The size of the voice frame (typically from 10 to 30 ms) and voice packet generation interval influence the number of packets in audio stream.
- If the mechanisms like *VAD/CNG/DTX* are used. Some of the speech codecs have those mechanisms built-in, for some of them they must be additionally implemented. These solutions influence the number of packets that are generated during VoIP call. The lower number of packets are transmitted the lower total covert channel bandwidth $B_T$ value.
- The *probability value of the packet loss* in IP network. Firstly, if this value is high we lose certain number of packets that are sent into the network, so the information covertly transferred within them is also lost. Secondly, while using delayed audio packets steganography we must adjust the probability of the intentionally lost packets to the level that exists inside the network to be sure that the perceived quality of the call is not degenerated.
- Less important steganographic methods specific conditions like: how often are RTCP reports are sent to the receiving party or if security mechanisms for communication are used.

To evaluate measures presented at the beginning of Section 3 the following test scenario, as depicted in Fig. 7, has been setup. Two SIP User Agents were used to make a call – the signalling messages were exchanged with SIP proxy and the audio streams flowed directly between endpoints. Moreover RTCP protocol was used to convey information about network performance. Audio was coded with ITU-T G.711 A-law PCM codec (20 ms of voice per packet, 160 bytes of payload). The ACD (Average Call Duration) for this experiment was chosen based on duration of the call for Skype service [21] and for other VoIP providers. In [7] results obtained that ACD for Skype is about 13 minutes, while VoIP providers typically uses a value between 7 and 11 minutes. That is why we chose ACD for the experiment at 9 minutes. There were 30 calls performed and all diagrams show average results.

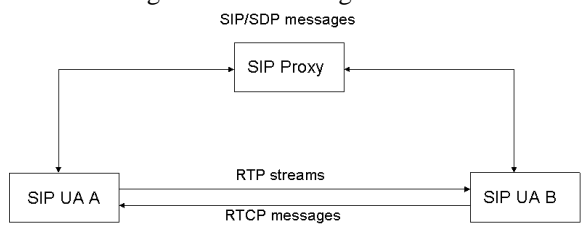

**Fig. 7.** VoIP steganography experimental test setup

The calls were initiated by *SIP UA A* and the incoming traffic was sniffed at *SIP UA B*. This way we were able to measure covert channel behavior for only one direction traffic flow. Based on the analysis of the available steganographic methods in section 3 the following steganographic techniques were used during the test (and the amount of data that were covertly transferred) as presented in Table 2.

**Table 2.** Steganographic methods used for experiment and their PRBR

| Steganographic method | Chosen PRBR |
|---|---|
| IP/UDP protocol steg. | 32 bits/packet |
| RTP protocol steg. | 16 bits/packet |
| RTCP steg. | 192 bits/packet |
| LACK | 1280 bits/packet (used 0.1% of all RTP packets) |
| QIM (audio watermarking) | 0.6 bits/packet |

We chose these steganographic methods for the experiment because they are easy to implement and/or they are our contribution. Besides they are the most natural choice for VoIP communication (based on the analysis' results from section 3) and, additionally, they represent different layers steganography. It is also important to note that assumed PRBR values for these methods were chosen to be reasonable in steganalysis context. We are interested however only in estimating a total amount of data that can be covertly transferred during the typical conversation phase of the VoIP call, and not how hard is to perform steganalysis. We want to see if the threat posed by steganography applied to VoIP is serious or not.

Achieved results of the experiment are presented below. First in Table 3 traffic flow characteristics, that were captured during performed VoIP calls are presented.

**Table 3.** Types of traffic distribution average results

| Type of traffic | Percent [%] |
|---|---|
| SIP messages | 0.016 |
| RTP packets | 99.899 |
| RTCP reports | 0.085 |

From Table 3 can be concluded that the steganographic methods that that utilizes RTP packets have the most impact on VoIP steganography as they cover 99.9% of the whole VoIP traffic. Next in Fig. 8 and Fig. 9 averaged results of the covert data flow distribution (RBR and PRBR respectively) during the average call are presented.

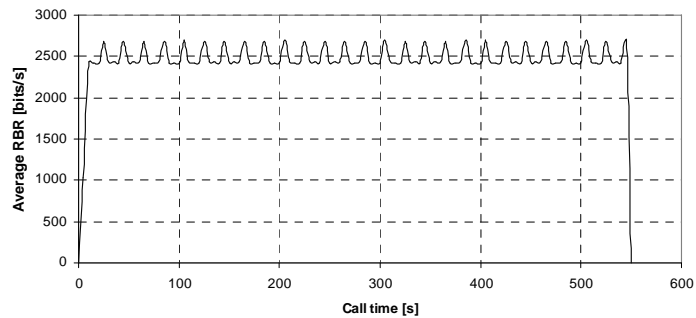

**Fig. 8.** Covert transmission data flow distribution for the experimental setup

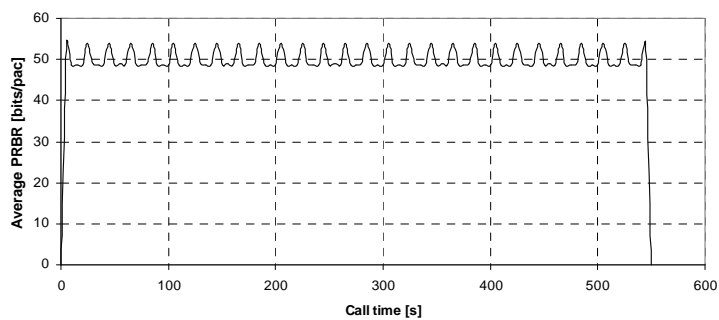

**Fig. 9.** PRBR during the average call

As one can see VoIP covert channels bandwidth expressed in RBR and PRBR changes in rather constant range during the call (between 2450 and 2600 bits/s for RBR and between 48 and 53 bits/packet for PRBR). The periodic peaks for curves presented in both figures are caused by steganographic bandwidth provided by LACK method. In every certain period of time packets are selected to be intentionally delayed and their payloads carry steganograms. For instants when these packets reach receiver the steganographic bandwidth increases. For this experiment the following average values were obtained and were presented in Table 4:

**Table 4.** Experimental results for typical call (for one direction flow only)

| Measure | Value | Standard Deviation |
| --- | --- | --- |
| Average total amount of covert data | 1364170 [bits] | 4018.711 |
| Average RBR | 2487,80 [bits/s] | 4.025 |
| Average PRBR | 50,04 [bits/packet] | 2.258 |

From the Table 4 we see that during the typical call one can transfer more than 1.3 Mbits (170 KB) of data in one direction with RBR value at about 2.5 kbit/s (50 bits/packet for PRBR).

**Table 5.** Types of traffic and theirs covert bandwidth fraction

| Type of traffic | Bandwidth fraction [%] | Bandwidth fraction [%] per steganographic method | |
| --- | --- | --- | --- |
| RTP packets | 99.646 | IP/UDP | 64.11 |
| | | RTP | 32.055 |
| | | Delayed audio packets | 2.633 |
| | | Audio watermarking | 1.202 |
| RTCP reports | 0.354 | - | |

As results from Table 5 show vast part of covert channels' bandwidth for VoIP is provided by network steganography (for protocols IP/UDP it is about 64% and for RTP 32%). Next steganographic method is delayed audio packets steganography (about 2.6%) and audio watermarking (about 1.2%). RTCP steganography provides only minor bandwidth if we compare it with other methods.

## 5  Conclusions

In this paper we have introduced two new steganographic methods: one of them is RTP and RTCP protocols steganography and the second is intentionally delayed audio packets steganography (LACK). We also briefly described other existing steganographic methods for VoIP streams. Next, for chosen steganographic method the experiment was performed. Obtained results showed that during typical VoIP call we are able to send covertly more than *1.3 Mbits* of data in one direction.

Moreover, the next conclusion is that the most important steganographic method in VoIP communication experiment is IP/UDP/RTP protocols steganography, while it provides over 96% of achieved covert bandwidth value. Other methods that contribute significantly are delayed audio packets steganography (about 2.6%) and audio watermarking techniques (about 1.2%).

Based on the achieved results we can conclude that total covert bandwidth for typical VoIP call is high and it is worth noting that not all steganographic methods were chosen to the experiment. Steganalysis may limit achieved bandwidth of the covert channels to some extent. But two things must be emphasized. Firstly, currently there is no documented active warden implementation thus there are no real counter measurements applied in IP networks so all the steganographic methods can be used for this moment. Secondly, analyzing each VoIP packet in active warden for every type of steganography described here can potentially lead to loss in quality due to additional delays – this would require further study in future. So, whether we treat VoIP covert channels as a potential threat to network security or as a mean to improve VoIP functionality we must accept the fact that the number of information that we can covertly transfer is significant.

## References

1. Ahsan, K., Kundur, D.: Practical Data Hiding in TCP/IP. In Proc. of: Workshop on Multimedia Security at ACM Multimedia '02, Juan-les-Pins, France (2002)
2. Baugher, M., McGrew, D., Naslund, M., Carrara, E., Norrman, K.: The Secure Real-time Transport Protocol (SRTP), IETF, RFC 3711 (2004)
3. Bender, W., Gruhl, D., Morimoto, N., Lu, A.: Techniques for Data Hiding. IBM. System Journal,. vol. 35, Nos. 3&4., 313-336 (1996)
4. Cuervo, F., Greene, N., Rayhan, A., Huitema, C., Rosen, B., Segers, J.: Megaco Protocol Version 1.0. IETF, RFC 3015, (2000)

5. Fisk, G., Fisk, M., Papadopoulos, C., Neil, J.: Eliminating Steganography in Internet Traffic with Active Wardens. In Proc. of: 5th International Workshop on Information Hiding, Lecture Notes in Computer Science, 2578, 18–35 (2002)
6. Giffin, J., Greenstadt, R., Litwack, P.: Covert Messaging Through TCP Timestamps. In Proc. of: Privacy Enhancing Technologies Workshop (PET), 194-208 (2002)
7. Guha, S., Daswani, N., Jain, R.: An Experimental Study of the Skype Peer-to-Peer VoIP System. In Proc. of: IPTPS – Sixth International Workshop on Peer-to-Peer Systems (2006)
8. ITU-T Recommendation H.323: Infrastructure of Audiovisual Services – Packet-Based Multimedia Communications Systems Version 6, ITU-T (2006)
9. Johnston, A., Donovan, S., Sparks, R., Cunningham, C., Summers, K.: Session Initiation Protocol (SIP) Basic Call Flow Examples. IETF, RFC 3665 (2003)
10. Korjik, V., Morales-Luna, G.: Information Hiding through Noisy Channels. In Proc. of: 4th International Information Hiding Workshop, Pittsburgh, PA, USA, 42-50, (2001)
11. Lucena, N., Lewandowski, G., Chapin, S.: Covert Channels in IPv6. In Proc. of: 5th Privacy En-hancing Technologies Workshop, Lecture Notes in Computer Science 3856, 147-166, (2005)
12. Mazurczyk, W., Kotulski, Z.: New Security and Control Protocol for VoIP Based on Steganography and Digital Watermarking. In Proc. of: IBIZA 2006, Kazimierz Dolny, Poland (2006)
13. Mazurczyk, W., Kotulski, Z.: New VoIP Traffic Security Scheme with Digital Watermarking. In Proc. of: SafeComp 2006, Springer-Verlag, Lecture Notes in Computer Science 4166, 170-181, (2006)
14. Mazurczyk, W., Szczypiorski, K.: Covert Channels in SIP for VoIP Signalling. In Proc. of: 4th International Conference on Global E-security 2008 (ICGeS 2008), London, United Kingdom, Communications in Computer and Information Science (CCIS) 12, Springer Verlag Berlin Heidelberg, 65-72 (2008)
15. Murdoch, S., Lewis, S.: Embedding Covert Channels into TCP/IP. Information Hiding (2005) 247-26
16. Na, S., Yoo, S.: Allowable Propagation Delay for VoIP Calls of Acceptable Quality. In Proc. of: First International Workshop, AISA 2002, Seoul, Korea, August 1-2, 2002, LNCS, Springer Berlin / Heidelberg, Volume 2402/2002, 469-480 (2002)
17. Petitcolas, F., Anderson, R., Kuhn, M.: Information Hiding – A Survey. IEEE Special Issue on Protection of Multimedia Content (1999)
18. Rosenberg, J., Schulzrinne, H., Camarillo, G., Johnston, A: SIP: Session Initiation Protocol. IETF, RFC 3261 (2002)
19. Schulzrinne, H., Casner, S., Frederick, R., Jacobson, V.: RTP: A Transport Protocol for Real-Time Applications, IETF, RFC 3550 (2003)
20. Servetto, S. D., Vetterli, M.: Communication Using Phantoms: Covert Channels in the Internet. In Proc. of IEEE International Symposium on Information Theory (2001)
21. Skype, http://www.skype.com
22. Szczypiorski, K.: HICCUPS: Hidden Communication System for Corrupted Networks. In Proc. of: ACS'2003, October 22-24, 2003 Międzyzdroje, Poland, pp.31-40 (2003)
23. Takahashi, T., Lee, W.: An Assessment of VoIP Covert Channel Threats. In Proc. of 3rd International Conference on Security and Privacy in Communication Networks (SecureComm'07), Nice, France (2007)
24. US Department of Defense – Department of Defense Trusted Computer System Evaluation Criteria, DOD 5200.28-STD (The Orange Book) (1985)
25. Zander, S., Armitage, G., Branch, P.: A Survey of Covert Channels and Countermeasures in Computer Network Protocols. IEEE Communications Surveys & Tutorials, 3rd Quarter 2007, Volume: 9, Issue: 3, pp. 44-57, ISSN: 1553-877X (2007)